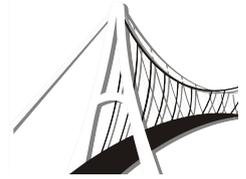

# Risk-based Design of Regular Plane Frames Subject to Damage by Abnormal Events: a Conceptual Study

André T. Beck, Lucas da Rosa Ribeiro, Marcos Valdebenito, Hector Jensen

# Risk-based Design of Regular Plane Frames Subject to Damage by Abnormal Events: a Conceptual Study

**André T. Beck[1]**

Department of Structural Engineering, University of São Paulo
Av. Trabalhador São-carlense, 400, 13566-590 São Carlos, SP, Brazil. atbeck@sc.usp.br

**Lucas da Rosa Ribeiro**

Department of Structural Engineering, University of São Paulo
Av. Trabalhador São-carlense, 400, 13566-590 São Carlos, SP, Brazil. lucasribeiro@usp.br

**Marcos Valdebenito**

Faculty of Engineering and Sciences, Universidad Adolfo Ibáñez.
Av. Padre Hurtado 750, 2562340 Viña del Mar, Chile. marcos.valdebenito.castillo@gmail.com

**Hector Jensen**

Departmento de Obras Civiles, Universidad Tecnica Federico Santa Maria
Av. España 1680, Valparaíso, Chile. hector.jensen@usm.cl

**ABSTRACT**

Constructed facilities should be robust with respect to loss of load-bearing elements due to abnormal events. Yet, strengthening structures to withstand such damage has a significant impact on construction costs. Strengthening costs should be justified by the threat and should result in smaller expected costs of progressive collapse. In regular frame structures, beams and columns compete for the strengthening budget. In this paper, we present a risk-based formulation to address the optimal design of regular plane frames under element loss conditions. We address the threat probabilities for which strengthening has better cost-benefit than usual design, for different frame configurations, and study the impacts of strengthening extent and cost. The risk-based optimization reveals optimum points of compromise between competing failure modes: local bending of beams, local crushing of columns, and global pancake collapse, for frames of different aspect ratio. The conceptual study is based on a simple analytical model for progressive collapse, but it provides relevant insight for the design and strengthening of real structures.

---

[1] Corresponding author

## 1. INTRODUCTION

Modern structural engineering requires built structures to be robust with respect to damage caused by abnormal events of exceptionally large intensity but low probability of occurrence. Following the recent partial or full collapses of buildings like Ronan Point Tower, Skyline Plaza, Alfred P. Murrah and the World Trade Center, design requirements for structural robustness were introduced in modern design codes (ASCE 7 2016; ASCE 41 2017; DoD 2013; GSA 2013). The design of robust structures is achieved by analyzing progressive collapse under potential initial damage.

Under multiple hazards, the probability of structural collapse can be evaluated as (Ellingwood and Dusenberry 2005; Ellingwood 2006, 2007):

$$p_C = P[C] = \sum_H \sum_{LD} P[C|LD, H] P[LD|H] P[H] \qquad (1)$$

where $C$ stands for collapse, $P[H]$ is the probability of hazard occurrence; $P[LD|H]$ is the conditional probability of local damage, given hazard $H$; $P[C|LD, H]$ is the conditional probability of collapse, given local damage $LD$ and hazard $H$. In Eq. (1), the sum over $H$ indicates the multiple hazards the structure is exposed to (for example, loads due to vehicular collisions, explosion, fire and terrorist attacks), and the sum over $LD$ represents the different initial damage states the structure can experience (local damage, support subsidence, internal or external column loss, penultimate column loss, etc.).

Some terms in Eq. (1) depend on structural mechanics, while others depend on human, social and political factors. A risk analysis of the structure, considering its surrounding environment and intended use, can address control of the hazards, or reduction of their rates of occurrence ($P[H]$). To some extent, protective measures arising from risk analysis can limit local damage produced by hazard $H$ (term $P[LD|H]$). Structural mechanics controls the terms $P[LD|H]$ and $P[C|LD, H]$; this last related to damage propagation following initial damage. Threat-independent approaches to robust design assume that local damage will occur, with loss of load bearing elements, and focuses on the damage propagation term.

The damage propagation analysis can be made independent of the non-structural (social, environmental, political) factors by considering initial damage probability as an independent parameter, following Beck (2020) and Beck *et al.* (2020):

$$p_{LD} = \sum_H P[LD|H] P[H] \qquad (2)$$

where $p_{LD}$ is the probability of local damage, like loss of a column, loss of a load-bearing wall, loss of a support, etc.

In this paper we study progressive collapse of regular frame structures subject to initial damage, like loss of columns and adjacent beams. A regular frame is understood as one with same bay length over height, and same height for all floors. Specifically, we address optimal design of regular frame buildings, considering the impacts of initial damage due to abnormal events. We employ the formulation of Beck *et al.* (2020), which considers the usual loading condition as one of the "hazards" in Eq. (1), with unitary probability of occurrence. The risk-based formulation looks for the minimum total expected costs of the building structure, which includes construction costs ($C_{constr.}$), cost of strengthening the frame to produce alternative load paths, cost of initial damage ($C_{LD}$), and eventual costs due to damage propagation ($C_{DP|LD}$) and/or collapse due to damage propagation ($C_{C|LD}$), where subscript $(\cdot)_{DP}$ is for damage propagation. The design variables are the design factors for beams ($\lambda_B$) and columns ($\lambda_C$). With these terms, the risk-optimization problem is stated as (Beck and Gomes 2012, Beck *et al.* 2015, Tessari *et al.* 2019):

$$\text{Find: } \{\lambda_B^*, \lambda_C^*\}$$
$$\text{which minimizes: } C_{TE}(\lambda_B, \lambda_C) = C_{constr.} + p_{LD}(C_{LD} + p_{DP|LD}C_{DP|LD} + p_{C|LD}C_{C|LD}) + p_f C_{NLC} \quad (3)$$
$$\text{subject to: } \lambda_B, \lambda_C > 0.$$

In Eq. (3), $p_f$ is the probability of failure under normal loading condition, and $C_{NLC}$ is the corresponding cost of failure term. This last term is considered mainly to cover the cases where $p_{LD}$ is very small, following Beck *et al.* (2020). Note that the probability of local damage $p_{LD}$ in Eq. (3) is the lifetime (herein, 50-year) probability. This can be related to yearly threat probabilities by: $p = -\ln[1 - p_{LD}]/50$.

The damage propagation and collapse terms in Eq. (3) depend on local failure of beams, local failure of columns adjacent to the initial damage, and global failure of columns (pancake failure). These probability and cost terms are described later in the paper.

In this paper, we investigate optimal designs resulting from Eq. (3), when applied to regular plane frames of varying aspect ratios, extent of initial damage, strengthening decisions and other factors. The analysis is a significant extension of results presented in Beck *et al.* (2020) and Beck *et al.* (2021). In particular, herein we address the competition between local bending, local crushing and global pancake failure modes, for frames of varying aspect ratios. In this paper, we do not address practical design aspects such as binding, structural fuses, compressive arch and Vierendeel actions. We employ a simple analytical model for progressive collapse, which considers plastic bending collapse of beams, and crushing collapse of columns. The model is limited to plane frames and to gravitational loads. Results provide insight that can be useful for actual structural design, but which needs to be verified using more complete models (Gerasimidis and Sideri 2016, Pantidis and Gerasimidis 2018) and specialized software (Adam *et al.* 2019).

One important aspect of design for robustness is the threat probability which justifies strengthening structural elements to provide alternative load paths, for instance. Addressing this issue, Beck *et al.* (2020) introduced the concept of a *threshold column loss probability*. Herein, we address local initial damage of different magnitude, potentially affecting a larger number of columns; also beams and slabs. Therefore, we recall the concept by giving it a more general name:

"*Local damage* probability threshold $p_{LD}^{th}$ is the value above which design for alternative (load) paths under discretionary local damage has positive cost-benefit, in comparison to usual design."

In this paper, we make an extensive investigation of how the $p_{LD}^{th}$ value changes for frames of varying aspect ratios, for different extents of initial damage and strengthening decisions. The remainder of this paper is organized as follows. The mechanical model for damage propagation and collapse of regular frames under gravity loads is presented in Section 2. Formulation of the cost functions for risk-based optimization is presented in Section 3. Numerical results are presented in Section 4, for a reference frame case, and in Section 5, for several variants. Conclusions are presented in Section 6.

## 2. PROGRESSIVE COLLAPSE OF REGULAR PLANE FRAMES: MECHANICAL MODEL

### 2.1 Basic formulation

This paper addresses design and strengthening of regular two-dimensional multi-story multi-bay frames, as illustrated in Figure 1. The mechanical model for damage propagation and progressive collapse, under gravity loads and initial damage, is based in Masoero *et al.* (2013). The model targets steel or Reinforced Concrete (RC) frames, and considers bending failure of beams, and crushing failure of columns. The model considers regular frames with $n_c$ columns and $n_s$ storeys, and an initial damage event leading to failure of $n_{r,c}$ columns and $n_{r,s}$ storeys, where subscript $r$ is for 'removed'. Here, $n_{r,s}$ refers to the vertical extent of the initial damage (Figure 1a). Beam length is $L$ for all spans, and column height is $H$ for all stories.

The simple analytical model of Masoero *et al.* (2013) has a few acknowledged limitations, which should be considered when interpreting results in this paper:

1. The model is limited to gravitational loads: wind, earthquake and other lateral loads are not considered; out-of-plumbness is not considered.

2. As a plane model, floor and other three-dimensional effects are ignored. Neglecting floor effects may underestimate the resistance significantly (He *et al.*, 2019).
3. Compressive arching effects significantly increase the resistance to progressive collapse, but are not considered in the model.
4. Strength and failure of beam-column connections is not considered.

Albeit simple, the model is useful for a conceptual cost-benefit analysis of design against progressive collapse. With removal of $n_{r,c}$ columns of $n_{r,s}$ stories, the frame may suffer bending collapse of $n_{r,c} + 1$ bays, or local crushing failure of two adjacent columns, as illustrated in Figure 1(b). Local crushing may propagate horizontally, leading to bending collapse of $n_{r,c} + 3$ bays, and so on, eventually extending the full horizontal extent of the frame, leading to global pancake collapse.

In the following, superscript $I$ is employed for the intact structures. Superscripts $B$ and $P$ refer to bending and pancake collapse. Pancake collapse can be local ($P, loc$) or global ($P, gl$). Furthermore, perfect brittleness is denoted $el$, whereas ideal plastic behavior is written as $pl$. This notation follows Masoero *et al.* (2013) for easy cross-referencing. The whole formulation is presented in terms of distributed loads $q_u$. The plastic hinge moment for beams is $B_y$, and crushing strength of columns is $R_c$.

For the intact structure, the ultimate strength in bending collapse is obtained as:

$$q_u^{I,B,pl} = \frac{16\,B_y}{L^2}. \tag{4}$$

This plastic solution is derived from the kinematic theorem, considering a triple-hinge plastic mechanism. In case of damage to $n_{r,c}$ columns, bending collapse strength is given by:

$$q_c^{B,pl}(n_{r,c}) = \frac{4\,B_y}{L^2(n_{r,c})}. \tag{5}$$

For the intact structure, static crushing of the columns occurs when maximum compressive force at base equals total axial strength, leading to the global pancake collapse strength:

$$q_u^{I,P} = \frac{R_c}{L\,n_s}\frac{n_c}{(n_c-1)}. \tag{6}$$

In case of local damage, local and global pancake collapses are possible. In case of local pancake, overload is directed to the two nearest intact columns. Brittle failure occurs for:

$$q_c^{P,loc,el}(n_{r,c}, n_{r,s}) = \frac{R_c}{L\,n_s}\frac{1}{\left(2 - \frac{n_c-1}{n_c} + n_{r,c}\left(1 - \frac{n_{r,s}}{n_s}\right)\right)}. \tag{7}$$

In case of damage to $n_{r,c}$ columns, brittle global pancake collapse occurs for:

$$q_c^{P,gl,el}(n_{r,c}, n_{r,s}) = \frac{R_c}{L\, n_s} \frac{n_c(n_c - n_{r,c})}{\left((n_c-1)(n_c+n_{r,c}) - 2\frac{n_{r,s}}{n_s} n_{r,c} n_c\right)}. \tag{8}$$

In our implementation, we consider ductile failure of beams, and brittle elastic failure of columns. Hence, our analysis is representative of RC frame buildings. Following the developments of Masoero et al. (2013), the dynamic load amplification factor is $DAF = 1$ for beams, and $DAF = 2$ for columns. This is closely related to the recommended $DAF$ values for deformation-controlled plastic failure of beams, and load-controlled failure of columns (GSA, 2013).

**2.2 Reference case**

Several frame and initial damage configurations are studied, as detailed later in the paper. In order to organize comparisons, a so-called ***reference*** case is initially considered: a frame with eight stories and eight bays ($n_s \times (n_c - 1) = 8 \times 8$), with beam length $L = 2H = 6\ m$. This frame is initially designed considering normal loading condition, and later strengthened considering discretionary removal of a single column and two adjacent beams from one bay ($n_{r,c}^0 = 1, n_{r,s}^0 = 1$). The superscript $(\cdot)^0$ refers to the initial discretionary damage for which the frame is strengthened.

The reference value for the probability of local damage is $p_{LD} = 0.1$, as detailed later. This fifty-year probability corresponds to an annual threat probability of $2.1 \times 10^{-3}$. Risk optimization results are computed for $10^{-6} \leq p_{LD} \leq 1$.

**2.3 Design under normal loading condition**

In the following, numerical results are presented for the *reference* frame configuration. The nominal dead and live loads are $L_n = D_n = 1\ (kN/m)$. Under normal loading condition, the required beam strength is (ASCE 7, 2016):

$$B_y^{NLC} = \frac{L^2}{16\phi} q_u^{I,B,pl} = \frac{L^2}{16\phi}(1.2 D_n + 1.6 L_n) = 7.41\ \text{kNm}. \tag{9}$$

Superscript $(\cdot)^{NLC}$ refers to the normal loading condition. The required column strength is:

$$R_c^{NLC} = \frac{L\, n_s(n_c-1)}{\phi n_c} q_u^{I,P} = \frac{L\, n_s(n_c-1)}{\phi n_c}(1.2 D_n + 1.6 L_n) = 140.55\ \text{kN}. \tag{10}$$

In Eqs. (9) and (10), $\phi = 0.85$ for a RC frame. In the remainder of this text, the frame designed under normal loading condition is referred to as ***normal*** frame.

## 2.4 Strengthening under initial damage condition

In the following, numerical results are presented for the reference case, where strengthening is considered for one column and two beams removed from one floor. Yet, it is assumed that any column of the first floor could be lost; hence, all columns and all beams of the first two floors need to be strengthened. The superscript $(\cdot)^0$ refers to the initial damage, for which the frames are strengthened. This is to distinguish from progressive failure involving damage to $n_{r,c}$ columns.

Beams and columns are strengthened in such a way as to provide an alternative path to the loads supported by the removed elements (Alternate Path Method, GSA 2013). For the beams to bridge over a single removed column, the required bending strength is (ASCE 7, 2016):

$$B_y^0(n_{r,c}^0 = 1) = \frac{L^2(n_{r,c}^0)}{4\phi} q_c^{B,pl} = \frac{L^2}{4\phi}(1.2D_n + 0.5L_n) = 15.3 \text{ kNm}. \tag{11}$$

Considering load redistribution, required strength for adjacent columns is:

$$R_c^0(n_{r,c}^0, n_{r,s}^0) = Max\left[R_c^{NLC}, \frac{L\, n_s}{\phi}\left(2 - \frac{n_c-1}{n_c} + n_{r,c}^0\left(1 - \frac{n_{r,s}^0}{n_s}\right)\right) q_c^{P,loc,el}\right] = 162.1 \text{ kN}. \tag{12}$$

with $q_c^{P,loc,el} = (1.2D_n + 0.5L_n) = 1.7$ kN/m. In strengthening, we consider $\phi = 1$, as recommended in ASCE 41 (2017). In Eq. (12), operator $Max[]$ is used to also consider the required column strength under normal loading condition. This operator is not required in Eq. (11) because the impact of a lost column in bending moments is large for $L \geq H$. The resulting strengthening factors $(\cdot)_{sf}$ for beams and columns are:

$$B_{sf} = B_y^0(n_{r,c}^0)/B_y^{NLC} = 2.06 \quad \text{and} \quad R_{sf} = R_c^0(n_{r,c}^0, n_{r,s}^0)/R_c^{NLC} = 1.23. \tag{13}$$

In order to address optimal design under an initial damage condition, we use independent design factors for beams ($\lambda_B$) and columns ($\lambda_C$). These are in addition to the usual code-recommended factors, such that $\lambda_B = \lambda_C = 1$ recovers the code-recommended design (ASCE 7 2016, ASCE 41 2017). Hence, strength equations (Eqs. 4 to 8) are computed with $B_y$ replaced by $B_y = \lambda_B B_y^0$, and with $R_c$ replaced by $R_c = \lambda_C R_c^0$, where $B_y^0$ and $R_c^0$ are the initial values of beam plastic hinge moment and column crushing capacities, respectively. In the remainder of this text, the frame designed under discretionary initial damage is referred to as **strengthened** frame. Strengthening two floors of the reference frame, as detailed above, has an impact of 13% on total construction costs, as detailed later.

## 2.5 Limit states and reliability analysis

The risk-based cost-benefit analysis addressed herein combines failure of the intact structure with failure/progressive collapse under initial damage conditions. For the intact structure, the limit state function for bending and global pancake collapse is:

$$g_I(\lambda_B, \lambda_C, \mathbf{X}) = R_I\, r_I(\lambda_B, \lambda_C,, \dots) - D - L_{50}, \tag{14}$$

where $r_I(\cdot)$ is a deterministic strength function for the intact structure, $R_I$ is a non-dimensional random variable describing uncertainty in strength of the intact structure, including model error, $D$ is the dead load, $L_{50}$ is the fifty-year extreme live load, and $\mathbf{X}$ is the vector of random system parameters. Usually, Eq. (14) is not a function of the progressive collapse design factors $\lambda_B$ and $\lambda_C$. However, as the structural elements are strengthened for load bridging under discretionary element removal, the reliability index for normal loading condition also becomes a function of $\lambda_B$ and $\lambda_C$. The strength function $r_I(\cdot)$ for bending failure is given by Eq. (4), and for global pancake by Eq. (6).

In case of localized initial damage, the limit state function is given by:

$$g_{LD}(\lambda_B, \lambda_C, \mathbf{X}) = R_D\, r_D(\lambda_B, \lambda_C, n_{r,c}, n_{r,s} \dots) - D - L_{apt}, \tag{15}$$

where $r_D(\cdot)$ is a deterministic strength function for the damaged structure, $R_D$ is a non-dimensional random variable describing uncertainty in strength of the damaged structure, including model error, and $L_{apt}$ is the sustained component of live load. In Eq. (15), the strength function $r_D(\cdot)$ for bending is Eq. (5), for local crushing is Eq. (7), and for global pancake failure is Eq. (8). Note that these equations are valid for any number of removed columns ($n_{r,c} < n_c$). This includes the initial triggering event, as well as progressive failure of columns.

By treating beam and column strength as the product of deterministic functions ($r_I$ and $r_D$) by single random variables ($R_I$ and $R_D$), the limit state functions become linear, allowing a second-moment solution. Statistics for $R_I$ and $R_D$ are different for beam and columns, as illustrated in Table 1. The statistics for these variables reflect material variability and model error, and were obtained by simulation from data in (Nowak *et al.* 2011). Data on random variables $R_I$ and $R_D$ should be reviewed in more practical applications, especially when considering horizontal loads and other failure models.

By approximating the probability distributions of $L_{apt}$ and $L_{50}$ as Gaussian, the limit state functions become linear functions of Gaussian variables, and reliability can be computed by the Cornell reliability index (Melchers and Beck 2018). This is considered accurate enough for the conceptual problem addresses herein. For more practical applications, use of FORM or Monte Carlo simulation is recommended (Ang and Tang 2006, Melchers and Beck 2018).

For an intact frame, the fifty-year Cornell reliability index is:

$$\beta^{50}(\lambda_B, \lambda_C) = \frac{r_I(\lambda_B, \lambda_C)\,\mu_{R_I} - (\mu_D + \mu_{L_{50}})}{\sqrt{r_I(\lambda_B, \lambda_C)^2 \sigma_{R_I}^2 + (\sigma_D^2 + \sigma_{L_{50}}^2)}}, \qquad (16)$$

where $\mu$ is the mean and $\sigma$ is the standard deviation of the corresponding variables. The reliability index for bending collapse of the intact frame is obtained for $r_I(\lambda_B) = q_u^{I,B,pl}$ (Eq. 4), with $B_y = \lambda_B B_y^0$; for global pancake collapse, $\beta^{50}$ is computed with $r_I(\lambda_C) = q_u^{I,P}$ (Eq. 6), with $R_c = \lambda_C R_c^0$.

Given local initial damage, the conditional arbitrary-point-in-time (*apt*) reliability index is:

$$\beta = \beta^{apt}(\lambda_B, \lambda_C, n_{r,c}, n_{r,s}) = \frac{r_D(\lambda_B, \lambda_C, n_{r,c}, n_{r,s})\,\mu_{R_D} - (\mu_D + \mu_{L_{apt}})}{\sqrt{r_D(\lambda_B, \lambda_C, n_{r,c}, n_{r,s})^2 \sigma_{R_D}^2 + (\sigma_D^2 + \sigma_{L_{apt}}^2)}}, \qquad (17)$$

The reliability indexes for bending collapse ($\beta_B$), local pancake ($\beta_{PL}$) and global pancake ($\beta_{PG}$) are obtained using Eqs. (5), (7) and (8), respectively. In the following, since we mainly refer to local damage condition, the superscripts $(\cdot)^{50}$ and $(\cdot)^{apt}$ are dropped for clarity of notation, when there is no risk of confusion. Reliability indexes obtained for the initial design of the *reference* frame are presented in Table 2. Numbers presented in gray color are those that are not usually computed, but which are presented here for completeness. Individual lines in Table 2 show how reliability changes along the design process: from the initial design, under normal loading condition (Eqs. 9, 10), to strengthening using Eqs. (11, 12) with $\lambda_B = \lambda_C = 1$, starting from the intact condition and moving to the conditional damage condition. The last column shows reliability indexes obtained from the risk optimization, as detailed in the sequence.

## 3. FORMULATION OF THE COST-BENEFIT RISK OPTIMIZATION PROBLEM

The decision to strengthen a structure, to make it able to bridge over a removed load-bearing element, has obvious impact on construction costs. In order to investigate the cost-benefit of different strengthening decisions, in potential initial damage scenarios, the costs of strengthening have to be confronted with the costs of building the structure, and the expected costs of failure. Costs of failure include costs of initial damage, cost of damage propagation and eventually, cost of full frame collapse.

## 3.1 Construction cost

In this paper, all cost terms are evaluated w.r.t. the cost for building the frame structure. Hence, cost of the structural frame is the reference cost, $C_{REF}$. Costs of structural materials vary significantly with geographical location, in terms of absolute values, and in terms of relative cost of steel to concrete. Cost of a RC structure, for instance, depends on cost of reinforcing steel, cost of concrete and cost of formwork and steel forming. These costs vary significantly due to, for example, geographical factors. As a simplification, and in benefit of generality, we assume unitary costs per length of beams and columns. For a span-to-height ratio of $L = 2H$, and for $L = 6$ m, the cost per meter of optimized beams and columns was found to be about the same by Boito and Kripka (2020). Hence, the reference cost is given by total linear length of the frame:

$$C_{REF} = L\, n_s(n_c - 1) + H\, n_s\, n_c. \tag{18}$$

This is understood as the cost for designing the frame under normal loading conditions. If the frame is strengthened to bridge over failed elements, construction costs increase. Typically, design for progressive collapse is of secondary nature (He *et al.*, 2019); hence, after main elements are sized considering normal loading condition (Eqs. 9, 10), they are verified under exceptional loading or element removal condition, and eventually strengthened. Typically, strengthening is done by increasing steel reinforcement area. When a column fails under a multi-span beam, maximum moments at the beam section above the column go from negative to positive. Hence, one immediate strengthening action is to use double instead of single reinforcement.

Considering typical ductile RC beams we found that, in order to double the strength of a beam in bending, it is required to roughly double the steel area. The impact in construction costs depends on the participation factor of steel to total costs, given as $\alpha_B$ for beams. Hence, the cost of strengthening the beams of each floor, for bridging over $n_{r,c}^0$ removed columns, is proportional to the following factor:

$$\left(\lambda_B \alpha_B B_{sf} + (1 - \alpha_B)\right). \tag{19}$$

Note that Eq. (19) includes the bending design factor $\lambda_B$. Using construction cost tables for Brazil (SINAPI, 2020), we found that participation of steel in total construction costs of a beam is roughly 70%. Hence, our reference value is $\alpha_B = 0.7$, but other values are also considered later in the paper. In Eq. (19), $(1 - \alpha_B)$ is the fixed part of beam construction costs.

Using similar reasoning, the cost factor for strengthening frame columns is written as:

$$\left(\lambda_C \alpha_C R_{sf} + (1 - \alpha_C)\right). \tag{20}$$

Finding the participation cost factor of columns is more difficult, as it strongly depends on concrete strength, load eccentricity and other factors. Hence, in order to reduce the number of parameters in the analysis, we consider $\alpha_C = \alpha_B = \alpha = 0.7$ in the following, unless stated otherwise. These factors should be verified when addressing more practical problems.

To keep the above cost terms in perspective, Praxedes and Yuan (2021a, 2021b) strengthened two four-story four-bay frames to bridge over loss of a single internal column. Their strengthening action was to increase steel ratio of about half the frames (covering beams and columns above two out of four bays). By comparing construction cost of the strengthened frames, with cost of the original frames, we arrive at $0.3 \lesssim \alpha \lesssim 0.4$. If the whole frames were strengthened, one would obtain $0.6 \lesssim \alpha \lesssim 0.8$.

Another important strengthening decision refers to the number of columns, and the number of beams to be strengthened. In a fully threat-independent design, all beams and columns should be strengthened. However, such as decision has strong impact in construction costs, and was shown not to be economical for typical threat probabilities (Beck *et al*. 2020, Praxedes 2020, Praxedes and Yuan 2021a, 2021b). For some threats like traffic accidents and explosions due to malevolent actions, first floor and building entrance columns are obvious targets. However, extent of the initial damage may not be limited to the first floor. Hence, in this paper the reference strengthening action is to reinforce all columns and all beams of the first two floors. This goes in line with the findings of Praxedes and Yuan (2021b) that optimal robustness is obtained when most reinforcement is allocated to the first floor, followed by the second floor.

In order to simplify notation, the unitary construction costs of beams and columns are written as:

$$C_{beams}(\lambda_B, n_{reinf,s}) = \{n_s - n_{reinf,s} + n_{reinf,s}(\lambda_B \alpha_B B_{sf} + (1 - \alpha_B))\}; \tag{21}$$

$$C_{columns}(\lambda_C, n_{reinf,s}) = \{n_s - n_{reinf,s} + n_{reinf,s}(\lambda_C \alpha_C R_{sf} + (1 - \alpha_C))\}. \tag{22}$$

The above terms include original cost of construction, plus cost of strengthening per unit length. The total construction cost is:

$$C_{const.}(\lambda_B, \lambda_C) = \frac{1}{C_{REF}} \left[ L\,(n_c - 1) C_{beams}(\lambda_B, n_{reinf,s}) + H\,n_c\,C_{columns}(\lambda_C, n_{reinf,s}) \right]. \tag{23}$$

For usual design under normal loading conditions, the same cost functions are considered, but with no strengthening ($n_{reinf,s} = 0$).

## 3.2 Cost of failure

For the conceptual study presented herein, cost of failure is assumed proportional to the extent of the damaged frame area. For an initial damage event leading to loss of $n_{r,c}^0$ columns from $n_{r,s}^0$ floors, the cost of initial damage is:

$$C_{ID} = \frac{1}{C_{REF}} \left( 2 L \, n_{r,s}^0 + H \, n_{r,c}^0 \right). \tag{24}$$

Failure consequences are strongly dependent on non-structural factors, such as building surrounding environment and intended use. Consequences of structural failure involve the costs of shutdown for rehabilitation and repair (lost revenue), costs for removing debris and rebuilding, damage to building contents and neighboring facilities, injury, death, and environmental damage. Out of these, only the cost of reconstruction depends on design safety margins. Hence, to separate non-structural consequence factors from the structural reliability analysis, as advocated in Beck (2020) and Beck *et al.* (2020, 2021), failure consequences are considered via an independent cost parameter $k$. Failure cost multipliers are a significant source of epistemic uncertainty, and they can change significantly for different structures, real estate market conditions, and economy interest rates (as failures occur in the future, w.r.t construction time).

Cost of construction in Eq. (23) is the cost of the structural frame. Marchand and Stevens (2015) studied ratios between costs of entire buildings and construction costs of structural frames. These ratios were found to be 6.8 for RC frames, 16.7 for steel frames, 4.4 for cold-formed steel and 10.5 for wood structures. Collapse failure costs are easily higher than the costs for reconstructing the whole building, and not just the structural frame. Hence, collapse failure cost multipliers can be significantly higher than the figures above; and should be considered constant, instead of functions of $\lambda_B$ and $\lambda_C$.

Financial losses from partial collapse of the Alfred P. Murrah Federal Building were estimated at $652 million by Hewitt (2003). This eight-story reinforced concrete building was built in 1977 at a cost of $14.5 million, or 24.7 million 1995 dollars at annual interest of 3%, which yields a total loss to building cost ratio of 26.4. The partial collapse affected 42% of the floor area of the building; a full collapse could bring this figure up significantly. Total losses arising from the 9/11 WTC attacks resulted 40 times larger than the building cost (Stewart 2017; Thöns and Stewart 2020); whereas a factor of 20 was found for the Pentagon (Muller and Stewart, 2011). In a cost-benefit analysis of terrorism risk-reduction measures for buildings, Stewart (2017) considered total loss to building cost ratios in the range 20-40. Considering the figures above as a reference, herein we consider $k = 40$ as a base case, and an upper-range value $k = 80$.

For engineering structures, brittle failures are usually more critical than ductile failures. Ductile failures provide warnings, allowing damaged structures to be evacuated, whereas brittle failures occur with little or no warning. When RC elements are properly designed, bending failure of beam/slabs is mostly ductile. The simple model by Masoero *et al.* (2013) does not consider column slenderness, nor bending-compression failure of columns. Since only axial load capacity is considered, we assume crushing failure of columns to be brittle. To account for the different consequences of failure, the cost multiplier for brittle failure (local and global pancake collapse) is twice that of ductile failure: $k_{brittle} = 2\, k_{ductile} = 40$, unless otherwise stated.

Costs of local collapse failures by bending or pancake are given by the impacted frame area (total linear length) immediately above the removed or failed column, multiplied by failure cost multipliers $k$. Cost of bending collapse is computed as:

$$C_B(n_{r,c}) = \frac{k_{ductile}}{C_{REF}} \Big( Min[n_{r,c} + 1, n_c - 1] L\, C_{beams}(\lambda_B = 1) + Min[n_{r,c}, n_c] H\, C_{columns}(\lambda_C = 1) \Big), \tag{25}$$

and cost of local pancake collapse is computed as:

$$C_{PL}(n_{r,c}) = \frac{k_{brittle}}{C_{REF}} \Big( Min[n_{r,c} + 3, n_c - 1] L\, C_{beams}(\lambda_B = 1) + Min[n_{r,c} + 2, n_c] H\, C_{columns}(\lambda_C = 1) \Big). \tag{26}$$

In Eqs. (25) and (26), the $Min[\,]$ operators warrant that costs of local collapse will not exceed costs of global collapse, if local pancake progresses into global pancake collapse. Evolution of cost of local failures, in terms of number of removed columns, is illustrated in Figure 12 of Beck *et al.* (2020).

Cost of global pancake collapse is given by frame volume, times failure cost multiplier:

$$C_{PG} = k_{brittle}\, C_{const.}(\lambda_B = 1, \lambda_C = 1). \tag{27}$$

Note that failure costs are computed w.r.t. the strengthened frame, but for unitary design factors $\lambda_B = \lambda_C = 1$. This makes the optimization problem more stable, according to our experience (Tessari *et al.* 2019, Beck 2020). This can be justified as damage to building contents, lost revenue and costs of compensation are one order of magnitude higher than structural material costs (as reflected by $k \geq 40$).

Equation (27) is also the collapse failure cost under normal loading conditions: since the frame is regular, bending failure of one beam, or crushing failure of one column, under uniform loading condition, represents simultaneous failure of all beams and all columns. This is an obvious simplification, as it neglects the bay-to-bay and story-to-story variations in loading and in member strengths, which are observed in practice.

### 3.3 Progressive collapse of the regular frame

If the frame suffers initial damage, leading to loss of $n_{r,c}^0$ columns and $n_{r,s}^0$ stories, progressive failure may occur. In the context of progressive failure, we refer to the number of failed columns ($n_{f,c}$), instead of number of removed columns ($n_{r,c}$). Clearly, the role of these variables in the formulation is the same, and the initial number of failed columns is $n_{f,c}^0 = n_{r,c}^0$. The chain of events that may follow initial damage includes:

1. Bending failure of beams of the $n_{f,c}^0 + 1$ affected bays, which may propagate upwards affecting all floors, but is otherwise self-arresting.
2. Local crushing failure of two adjacent columns, which may propagate horizontally, affecting two additional bays, with $n_{f,c} = n_{f,c}^0 + 2$, and so on, until it is naturally arrested, or until complete (global pancake) collapse.
3. Local crushing failure of $n_{f,c} = n_{f,c}^0 + 2$ columns may also cause bending failure of $n_{f,c}^0 + 3$ bays, and so on.

The likelihood of occurrence of the above progressive failure events is controlled by beam and column strengths, which depend on design factors $\lambda_B$ and $\lambda_C$. The conditional probabilities of occurrence of local bending, local pancake and global pancake collapse events are given as:

$$p_B = \Phi[-\beta_B(\lambda_B, n_{f,c})],$$
$$p_{PL} = \Phi[-\beta_{PL}(\lambda_C, n_{f,c})],$$
$$p_{PG} = \Phi[-\beta_{PG}(\lambda_C, n_{f,c})]. \tag{28}$$

In this problem, failure consequences are related to frame areas which overlap (see Figure 2 in Masoero *et al.* (2013)). For the same number of removed columns, the area affected by local pancake collapse includes the area affected by bending collapse. Progressive collapse due to local pancake, for $n_{f,c} + 2$, encompasses the same area affected by local pancake with $n_{f,c}$ failed columns. Global pancake failure affects the whole frame. In this setting, a reasonable approximation to the failure tree is to consider the maximum expected cost, amongst all possible failure events. This approximation is also possible because the failure modes are likely to be strongly correlated, as they depend on the same loads and initial damage event.

For the initial discretionary column removal event, the maximum expected cost becomes:

$$Max[p_B\ C_B(n_{f,c}^0),\ p_{PL}C_{PL}(n_{f,c}^0), p_{PG}C_{PG}(n_{f,c}^0)]. \tag{29}$$

Note that $C_{PG} > C_{PL} > C_B$, but the above terms are balanced by the corresponding probabilities, which depend on partial factors $\lambda_B$ and $\lambda_C$.

Local pancake collapse, with removal of $n_{f,c}^0$ columns, may evolve into bending collapse or progressive local pancake collapse, affecting $n_{f,c}^0 + 2$ columns, and so on. The conditional probability that local pancake collapse will advance, to involve $n_{f,c} + 2$ columns, is given by $p_{PL}(n_{f,c} + 2)$. The unconditional probability is: $p_{PL}(n_{f,c})p_{PL}(n_{f,c} + 2)$. Thus, the expected collapse cost for progressive failure is given by:

$$p_{PL}(n_{f,c})p_{PL}(n_{f,c} + 2)Max[p_B\ C_B(n_{f,c} + 2),\ C_{PL}(n_{f,c} + 2), p_{PG}C_{PG}(n_{f,c} + 2)]. \tag{30}$$

With these preliminaries, the total expected cost, for progressive collapse failure of the regular 2D frame, is given by:

$$\begin{aligned}
C_{TE}(\lambda_B, \lambda_C) &= C_{const.}(\lambda_B, \lambda_C) &\text{(a)}\\
&\quad + C_{const.}(1,1)\ \Phi[-\beta_B^{50}(\lambda_B)] + C_{PG}\ \Phi[-\beta_{PG}^{50}(\lambda_C)] &\text{(b)}\\
&\quad + p_{LD}\ Max[p_B\ C_B(n_{f,c}^0),\ p_{PL}C_{PL}(n_{f,c}^0), p_{PG}C_{PG}(n_{f,c}^0), \dots &\text{(c)}\\
&\quad \dots Max_{j=(n_{f,c}^0+2)}^{(n_c-2)}\{p_{PL}(j-2)p_{PL}(j)Max[p_B\ C_B(j),\ C_{PL}(j), p_{PG}C_{PG}(j)]\}\ ]. &\text{(d)}
\end{aligned}$$

$$\tag{31}$$

In Eq. (31), line (b) corresponds to global failure of the intact frame, under normal loading condition. Line (c) corresponds to the maximum expected cost in the initial discretionary local damage event. Line (d) accounts for the maximum expected cost during eventual progressive collapse. Note that the operator $Max[\ ]$ in line (c) extends to line (d), i.e., only the event leading to maximum expected failure cost is computed. This warrants that cost of collapse will not exceed $k_{brittle}\ C_{const.}(1,1)$. The counter $j$ in line (d) should vary in steps of two units.

### 3.4 Objective function for cost-benefit optimization

The cost terms defined in the last section, and grouped in Eq. (31), already include the expected cost of failure of the intact structure (due to bending or global pancake collapse), the expected costs of progressive failure, and the expected costs of collapse. Hence, the risk optimization problem in Eq. (3) is solved, considering the total expected costs in Eq. (31) as the objective function.

## 4. RESULTS FOR THE REFERENCE CASE

The formulation just presented was implemented in Mathematica 12. The optimization problem is solved using various built-in algorithms of function *Minimize* (Wolfram Research 2018). Results for the reference case are presented in this section.

### 4.1 Conditional failure probabilities, conditional and expected costs of failure

We start by illustrating the conditional failure probabilities in Eq. (28), the conditional costs of failure in in Eqs. (29) and (30), and the expected costs of failure in lines (b), (c) and (d) of Eq. (31).

Figure 2 a) to d) shows conditional failure probabilities, conditional costs of failure, progressive damage probabilities and expected costs of failure, as the number of failed columns increases, for the *reference* frame. The strong lines in these figures are the results for the *strengthened* frame, whereas the fading lines correspond to the *normal* frame (normal loading conditions). As expected, response of the strengthened frame, given initial damage, is better: failure probabilities are smaller for all failure modes (Figure 2a), and expected costs of failure are significantly smaller (Figure 2d).

In Figure 2a) we observe that conditional failure probabilities are very large for the normal frame, especially in bending. For the strengthened frame, conditional failure probabilities start smaller, but increase rapidly as the number of failed columns increases from one to three. Conditional probability of global pancake failure is small up until $n_{f,c} = 3$, and increases rapidly for $n_{f,c} > 3$. Conditional costs of failure (Figure 2b) are larger for the *normal* frame, for the initial number of removed columns ($n_{r,c}^0 = 1$). However, as the number of failed columns increases, these conditional cost terms become larger for the strengthened frame, since the total cost of the strengthened frame is larger than the total cost of the normal frame. Damage probabilities (Figure 2c) start at $p[LD|n_{f,c}] = 1$, for $n_{f,c} = 1$, but drop faster for the strengthened frame, as $n_{f,c}$ increases. Two opposing factors explain the V-shapes in expected costs of failure (Figure 2d): as the number of failed columns increase, the damage area increases, but the event probabilities decrease. As a result, we see that expected damage for the strengthened frame reaches a plateau which is much lower than the expected damage for the normal frame.

The conditional probabilities and cost terms illustrated in Figure 2 are basically the same terms of the Robustness Index recently proposed by Praxedes et al. (2021).

## 4.2 Optimal design of the reference frame

By solving the cost-benefit optimization problem in Eq. (3) for $p_{LD} = 0.1$, the optimal design values $\lambda_B^* = 0.9$ and $\lambda_C^* = 1.3$ are found. Hence, the optimizer reduces strength of beams, and increases strength of columns, in comparison to the strengthening resulting from Eqs. (11) and (12). Conditional failure probabilities, conditional costs of failure, local damage probabilities and expected costs of failure of the *optimally strengthened* frame are compared with the *strengthened* frame in Figure 3 a) to d).

As observed in Figure 3a), by reducing $\lambda_B$, the optimizer increases conditional bending failure probabilities for one and two lost columns. At the same time, the optimal solution reduces conditional local pancake probabilities for up to five lost columns. Global pancake collapse probability is reduced between 3 and 6 lost columns. Figure 3b) shows that the conditional cost of bending failure is not affected by the slight increase in conditional bending failure probabilities. Yet, the conditional costs of local and global pancake collapse are reduced, for the optimally strengthened frame. The changes observed in Figures 3 a) and b) are not very large, but they are significant in terms of reducing probability of damage propagation (Figure 3c) and total expected costs of failure (Figure 3d). Hence, the risk-optimization results in a better balance between the failure modes and the corresponding expected costs of failure.

## 5. RESULTS FOR OTHER FRAME CONFIGURATIONS

### 5.1 Problem variants

Several problem variants are considered in the sequence. This includes frames of different aspect ratios (number of stories × number of bays), as detailed in Table 3, as well as aspect ratio of the individual bays, failure cost multipliers, strengthening costs and size of initial damage (Table 4). Table 3 includes a tall frame with 16 stories and 4 bays, the reference "square" frame with 8 stories and 8 bays, a low frame 4 stories heigh with 16 bays, as well as intermediate cases, all with similar "tributary" area. In the sequence, the seven frame variations detailed in Table 3 are combined with the variations listed in Table 4. Further details about Table 4 variations are discussed with the results.

### 5.2 Optimal safety factors versus local damage probability

Figure 4 shows the optimal design factors $\lambda_B^*$ and $\lambda_C^*$ for tall and low buildings (Table 3), as a function of local damage probability ($p_{LD}$). Figure 5 shows corresponding optimal values of bending and pancake reliability indexes, also for tall and low frames. Results for the *reference* frame follow the same pattern and would be situated between those of the tall and the low frames in Figures 4 and 5. As can be observed, optimal reliability indexes (Fig. 5) follow the same overall trend of the optimal design factors (Fig. 4), although the relationship between them is not linear.

As observed in Figure 4, optimal design factors change significantly with the probability of initial damage; the only exception is the column design factor, which is indifferent to $p_{LD}$ for the low frame. For the tall frame, optimal values of $\lambda_C^*$ are as high as 1.6 for $p_{LD} = 1$, dropping to around 1.2 for smaller $p_{LD}$. Clearly, for tall frames with smaller number of columns, loss of a single column has much greater impact. Interestingly, for small $p_{LD}$, optimal $\lambda_C^*$'s for tall frames is smaller than for low frames, a result that may be counterintuitive in first sight. However, note that the overload or strengthening factor for adjacent columns, given loss of a single column, is $R_{sf} = 1.0$ for the low frame, but $R_{sf} = 1.38$ for the tall frame (Table 5). Hence, it is cheaper to strengthen columns of the low frame, as the additional design margin is (only) $\lambda_C^* \approx 1.4$ for these columns. For the tall frame, the cost of strengthening adjacent columns is proportional to $1.6 \times 1.38 = 2.2$, for $p_{LD} = 1$, dropping to around $1.2 \times 1.38 = 1.67$ for smaller $p_{LD}$. It is also relevant that for low frames, global pancake failure is unlikely for a single column failure; yet, the whole frame may collapse if local pancake failure progresses horizontally (this could also be avoided by structural fuses, which is not addressed herein). Hence, it is relatively cheaper to protect the low frame against progressive local pancake failure. For the tall frames, loss of a single column has greater impact in local and global pancake failure probabilities, as observed in Fig. 5.

Optimal design factors for bending are larger than one for tall frames and $p_{LD} \gtrsim 0.1$, and smaller than one otherwise. Optimal $\lambda_B^*$'s are significantly larger for tall buildings, because bending failures due to column loss propagate upwards, causing greater consequences for taller buildings. Optimal design factors for bending become significantly smaller as $p_{LD}$ is reduced. Very small values of $\lambda_B^*$ do not have practical significance, as minimal required beam strength would likely be determined by serviceability (displacement) limit states. Comparing Figures 4 and 5 it can be observed that, when $\lambda_B^*$ drops to about 0.6, the optimal bending reliability index $\beta_B^*$ drops to zero. As argued in Beck et al. (2020), this corresponds to an optimal design which is indifferent to objective consideration of discretionary column removal. This is detailed in the sequence.

### 5.3 Threshold local damage probabilities

As illustrated in Figures 4 and 5, the probability that a regular frame suffers local damage, leading to loss of one column and two beams of a single floor, has significant impact in optimal structural design. As $p_{LD}$ becomes smaller, the optimal design changes from an alternative load path, or load bridging design, with large design factors, to an alternative path design with minimal design factors, which eventually approaches the usual design (with no discretionary element removals). This becomes a smooth transition because the formulation proposed in Beck *et al.* (2020), and employed herein (Eq. 3), combines normal and abnormal loading conditions in the same objective function (Eq. 31).

As argued by Beck *et al.* (2020, 2021), a threshold local damage probability ($p_{LD}^{th}$) can be identified, above which design or strengthening considering discretionary element removals has better cost-benefit than design under normal loading conditions. Two different situations have to be acknowledged in this context: A) design and strengthening considering current normative (with $\lambda_B = \lambda_C = 1$); and B) designs resulting from risk optimization (Eqs. 3 and 31), with optimal values $\lambda_B^*$ and $\lambda_C^*$. The resulting $p_{LD}^{th}$'s are conceptually the same, but numerically different.

When current normative is considered (situation A), threshold local damage probabilities ($p_{LD}^{th}$) are identified by comparing total expected costs for usual design (Eqs. 9, 10), and for design/strengthening considering discretionary element removals (Eqs. 11, 12 with $\lambda_B = \lambda_C = 1$). This is illustrated in Figures 4 and 7 of Beck *et al.* (2020), and in Figures 4, 7 and 8 of Beck *et al.* (2021). This goes inline with the "practical" definition presented in Beck *et al.* (2020) and reproduced in the introduction. In this paper we address a variety of frames, with different aspect ratios, different initial damage and different strengthening measures. For some of the taller frames, conventional progressive collapse design with $\lambda_B = \lambda_C = 1$ is always cheaper than design under normal loading condition. For some of the lower frames, the opposite is true. For these cases, it is impossible to find a root of the difference between total expected costs; hence, the most practical interpretation of $p_{LD}^{th}$ cannot be employed.

For the risk optimization problem, the threshold local damage probability ($p_{LD}^{th}$) is a point of indifference of the optimal solution, where two local minima with similar objective function values are observed: one is an alternative path solution with reduced optimal design margin for bending failure ($\lambda_B^* < 1$); the other is a solution with $\lambda_B^* << 1$, which approaches the design under normal loading condition. These local minima solutions are illustrated in Figures 3, 8-9, 13-16 of Beck *et al.* (2020), and in Figures 3 and 9 of Beck *et al.* (2021). Local minima do not always exist, as seen in Figure 6 of Beck *et al.* (2020). Automatic identification of local minima is difficult to implement.

The indifferent behavior, leading to local minima of similar total expected costs, is associated with a transition, from positive to negative, of the optimal reliability index for bending, $\beta_B^*$. This transition is illustrated in Figure 5 and Table 5 of Beck *et al.* (2020), in Figure 5 of Beck *et al.* (2021), and can also be observed in Figure 5, for the tall and low frames considered herein. Hence, a practical way of identifying the indifferent design is by finding the $p_{LD}$ root for which the optimal bending reliability index $\beta_B^*$ is zero. This approach is adopted in this paper. The *Solve* function of Mathematica (Wolfram Research 2018) is used for root finding. As can be observed in Figure 5, for the low frame $p_{LD}^{th} \approx 0.05$, and for the tall frame, $p_{LD}^{th} \approx 10^{-3}$. Figure 6 shows $p_{LD}^{th}$ values for frames of different aspect ratio and following the problem variants in Tables 3 and 4.

As observed in Figure 6a), lower frames require higher local damage probabilities to justify strengthening with discretionary element removal. For taller frames, the threshold $p_{LD}^{th}$ values are smaller, dropping to about $10^{-4}$. This corresponds to an annual threat probability of $2 \times 10^{-6}$, which is of the order of magnitude of usual threats (gas explosions or fire, with $10^{-6}$ to $10^{-5}$ occurrences per year). This shows that design considering discretionary element removals can be economically justified for taller frames, in an extension to results presented in Beck *et al.* (2020).

To keep the numbers in Figure 6 in perspective, Thöns and Stewart (2019) found that protective measures for iconic bridges are not economical for threat probabilities smaller than $10^{-4}$ per year, and Stewart (2017) found that strengthening buildings is only cost-effective for threat probabilities larger than $10^{-3}$ per building per year. These numbers correspond to 50-year local damage probabilities between $5 \times 10^{-3} \leq p_{LD}^{th} \leq 5 \times 10^{-2}$, well within the range of $p_{LD}^{th}$ values in Figure 6.

### 5.4 Results for Set 1: effect of frame and bay aspect ratios
**Threshold probabilities**

Figure 6a) also shows threshold local damage probability results for bays of different aspect ratio. The purple line is for the *reference* case, with $L = 2H$. When bay length is reduced by half ($L = H$, dotted line), with corresponding reduction in tributary area ($A_U$), $p_{LD}^{th}$ values are reduced. When bay length is increased by 50% ($L = 3H$, dashed blue line), with corresponding increase in tributary area, $p_{LD}^{th}$ values are increased. Yet, if bay length is reduced by half, by doubling the number of columns ($L = H$, dash-dotted red line), a different behavior is observed: $p_{LD}^{th}$ values increase for lower frames but decrease for taller frames. Hence, we observe that the frame aspect ratio, as given by number of stories × bays, or $(n_s \times (n_c - 1))$, has greater impact on results than the actual aspect ratio of individual bays.

**Optimal design factors and reliability indexes for $p_{LD} = 0.1$**

As observed in Figures 4 and 5, optimal design factors and reliability indexes vary significantly with the initial local damage probability. In the following, we analyze how these optimal values change for frame of different aspect ratio, by fixing $p_{LD} = 0.1$. Recall that this 50-year probability corresponds to an annual threat probability of $2.1 \times 10^{-3}$. This value is above usual threat probabilities for column loss in buildings, but it is in the range for which progressive collapse design is cost-effective for all frames studied herein, as shown in Figure 6.

Figure 7 illustrates optimal design factors for beams ($\lambda_B^*$) and columns ($\lambda_C^*$), for all frame and bay aspect ratios considered herein. Optimal bending design factors are nearly unitary for tall frames and reduce continuously for lower frames. This matches the behavior observed in Figure 4: larger design factors are justified for taller frames, because bending failures would progress upwards. As frame height is reduced, the $p_{LD}^{th}$ values observed in Figure 6a) get closer to the fixed $p_{LD} = 0.1$ of Figure 7; this also explains why optimal $\lambda_B^*$'s are reduced. The smallest $\lambda_B^*$'s are obtained when the number of columns is doubled, in comparison to the *reference* case.

An interesting non-proportional effect is observed for the optimal column design factors. As frame height increases, optimal $\lambda_C^*$'s are reduced. This trend was observed in Figure 4: for lower frames, the column strengthening factor $R_{sf}$ is small, and the importance of avoiding (progressive) local pancake failures is large. Yet, for the three taller frames in Figure 7 this tendency is reverted, and optimal $\lambda_C^*$'s increase. This may be to avoid global pancake failures. Optimal $\lambda_C^*$'s for frames with doubled number of columns show a distinct, more indifferent behavior.

By looking to the joint behavior of $\lambda_B^*$'s and $\lambda_C^*$'s, and disconsidering the case with additional columns (dash-dotted red line), it is observed that for frames lower than $(11 \times 6)$, reductions of $\lambda_B^*$'s are accompanied by increases in $\lambda_C^*$. As the frames become lower and wider, the consequences of beam failures are reduced (upward progression), but the consequences of local pancake failures increase (horizontal progression). For frames taller than $(11 \times 6)$, this tendency changes, and both optimal design factors increase with increase in frame height (and reduction in number of columns). This occurs because consequences of beam and column failures increase with frame height. Hence, what we observe in Figure 7 is a competition between beam bending and column crushing failure modes. The resources allocated into frame strengthening need to be compensated by reductions in expected costs of failure. The optimal allocation of these resources, between beams and columns, changes according to the frame and bay aspect ratios, as observed in Figure 7. Actual values of optimal design factors in Figure 7 are valid for this paper only, but the identified trends should be valid for real structures as well.

Figures 8, 9 and 10 illustrate the optimal bending, local pancake and global pancake reliability indexes, respectively. Overall, the trends observed in Figure 7 can be identified in Figures 8 to 10. The fading lines in Figures 8 to 10 illustrate reliability indexes obtained for the usual progressive collapse design, with $\lambda_B = \lambda_C = 1$. For bending (Figure 8), usual design leads to constant $\beta_B$'s around $\beta_B = 2$; whereas optimal design has a large impact on bending failure probabilities. For local pancake (Figure 9) and global pancake (Figure 10), usual progressive collapse design leads to nearly constant collapse probabilities. These are significantly reduced by optimal design. Overall, it is observed in Figures 8 to 10 that optimal design finds a better balance between the different failure modes of the regular frame subject to loss of load-bearing elements.

## 5.5 Results for Set 2: effect of cost multipliers

It is expected that results of risk optimization depend, to a great extent, on failure cost multipliers $k$. Consequences of failure will vary significant with building use and occupancy, as well as surrounding environment. Figure 6b) illustrates threshold local damage probabilities $p_{LD}^{th}$ for different failure cost multipliers. The continuous purple line is the result for the *reference* case, with $k_{brittle} = 2\, k_{ductile} = 40$. As observed, increased failure consequences lead to a drop in $p_{LD}^{th}$ values, making strengthening for load bridging cost-effective for a greater range of frame structures. Changes in $k_{ductile}$ have a greater impact on $p_{LD}^{th}$, as this value is determined directly from the root $\beta_B^* = 0$ (ductile failure of beams is assumed). A doubling of $k_{brittle}$, for fixed $k_{ductile}$, produced minor impact on results (dotted and dash-dotted lines). Results for optimal design factors and reliability indexes are similar to those shown in Figures 7 to 10, in terms of relative trends. Overall, for larger cost multipliers, hence larger consequences of failure, optimal design factors and reliability indexes are larger. Optimal $\lambda_B^*$'s increase with $k_{ductile}$, and optimal $\lambda_C^*$'s increase with $k_{brittle}$.

## 5.6 Results for Set 3: effect of strengthening cost

In Beck *et al.* (2020) it was shown, for a single frame of 11 bays by 11 stories, that the decision to strengthen a structure for load bridging over failed load-bearing elements depends on strengthening costs. As discussed in Section 3.1, strengthening costs depend on local costs of materials, both absolute and relative. Strengthening costs also depend strongly on the strengthening decisions, for instance, the decision on the number of stories and bays to reinforce. Herein, the default strengthening decision involves all columns and all beams of the first two floors ($n_{reinf,s} = 2$). Herein, we investigate the effects of strengthening costs on the optimal design of a wider range of frames.

Figure 6c) illustrates local damage probability thresholds, $p_{LD}^{th}$, for different participation factors $\alpha_B$ and $\alpha_C$. Recall that $\alpha_B$ is the participation of cost of steel in strengthening RC beams: doubling the plastic hinge strength of beams requires doubling the amount of steel (approximately), and this would lead to a $2\alpha_B$ impact on the cost of strengthened beams. The continuous purple line in Figure 6c) is the *reference* case, with $\alpha_B = \alpha_C = 0.7$, and $n_{reinf,s} = 2$. When the cost participation factors increase to $\alpha_B = \alpha_C = 0.9$, threshold $p_{LD}^{th}$ values increase for all frames, making progressive design cost-effective only for larger threat probabilities. If $\alpha_B$ is reduced to $\alpha_B = 0.5$, with $\alpha_C = 0.9$, we observe that the impact of $\alpha_B$ reduction is larger than the impact of $\alpha_C$ increase (from 0.7). This is a direct consequence of evaluating $p_{LD}^{th}$ from the root $\beta_B^* = 0$. If participation factors are maintained at $\alpha_B = \alpha_C = 0.7$, but the decision is to reinforce the whole frames, $p_{LD}^{th}$ increases significantly for all frames. Hence, since the strengthening decision has greater impact on construction costs, it is justified only for larger threat probabilities. The treat probabilities which justify strengthening the whole frame are significantly larger than usual values for gas explosion or fire threats.

Figure 11 illustrates the optimal safety factors for beam bending and column crushing, in terms of the strengthening cost factors. The optimal $\lambda_B^*$'s vary in proportion to the beam cost participation factor $\alpha_B$, with larger $\lambda_B^*$'s obtained for smaller $\alpha_B$. Similar behavior is observed when $\alpha_C$ is increased: increasing $\alpha_C$ leads to a reduction in $\lambda_C^*$'s. When the relative value of cost factors changes, an unexpected behavior is observed: reducing $\alpha_B$, while keeping $\alpha_C$ constant, produced a reduction in optimal $\lambda_C^*$'s for lower frames; but an increase in $\lambda_C^*$'s for taller frames. For taller frames, the reduction in beam strengthening cost led to an increase in optimal $\lambda_B^*$'s, but also to an increase in optimal $\lambda_C^*$'s! This confirms the competition between failure modes, observed in Figure 7: for tall frames, stronger beams need to the accompanied by stronger columns, in order to avoid pancake failures. The dash-dotted red line in Figure 11 shows that the competition between failure modes is affected by the relative value of cost factors for strengthening beams and columns.

### 5.7 Results for Set 4: effect of extent of initial damage

One of the largest unknowns in alternative path (APM) design is the extent of initial damage, for which alternative load paths should be developed. This uncertainty relates to the actual response of the structure, given unknown initial damage, but also to the discretionary element removal scenarios for which APM strengthening is performed. To simplify matters, in this paper a match is considered between the design and the actual initial damage scenario. Mismatches should be addressed in future research.

Clearly, strengthening frames to sustain larger initial damage has an impact on construction costs. Table 5 shows the factors required for strengthening beams and columns to sustain the initial damages listed in Set 4 of Table 4 (and in the second line of Table 5). For the beams to sustain loss of 1, 2 and 3 columns requires plastic hinge strengths which are about 2 ×, 4 × and 6 × larger than the strength under normal loading condition. The strengthening factor for columns varies significantly with frame height, as shown in Table 5.

Figure 6d) shows how the local damage probability thresholds, $p_{LD}^{th}$, changes for different extents of initial damage. The *reference* case, with $(n_{r,c}^0 \times n_{r,s}^0) = (1 \times 1)$, is shown as a continuous purple line. As observed, a reduction in the number of affected beams ($n_{r,s}^0 = 0$) has no impact on $p_{LD}^{th}$ (lines are superimposed). Yet, increasing the number of removed columns ($n_{r,c}^0$) has a large impact, making APM design economical only for larger threat probabilities. Typically, the probability of initial damage is inversely proportional to the extent of initial damage. A comprehensive analysis would require addressing conditional probabilities of progressive collapse given one, two, or more removed columns. This will be addressed in future research. Figure 6d) shows an intermingling effect of the results for two and three columns removed; this also deserves further investigation.

Figure 12 illustrates the optimal safety factors for beam bending and column crushing, in terms of extent of initial damage. As observed, the initial damage has larger impact on optimal design factors for columns. Although load bridging over a larger span has significant impact in beam strengthening factors ($B_{sf}$ in Table 5), this does not reflect in large changes in optimal $\lambda_B^*$'s.

The competition between failure modes is also significantly affected by the extent of initial damage. The opposing trend between optimal $\lambda_B^*$'s and $\lambda_C^*$'s, which in Figure 11 was observed for frames lower than $(11 \times 6)$, is now observed only for the two lowest frames (right in Figure 12). With larger initial damage ($n_{r,c}^0 = 2$ or 3), the increase in $\lambda_B^*$ for taller frames is accompanied by increase in $\lambda_C^*$. For the larger extent of damage, strengthening beams makes pancake failures more likely. To avoid this, beams and columns need to be strengthened simultaneously.

## 6. EFFECTS OF CATENARY ACTION

Catenary action provides significant additional strength to beams in progressive collapse. So far in this paper, catenary effects have not been included, mainly because catenary effects are not objectively considered when determining required beam strengths and dimensions.

Equations (4) and (5) provide the distributed loads which produce plastic hinge mechanisms in beams of intact and damaged frames. Following Masoero et al. (2013), axial catenary effects can be considered by adding the following terms to Eqs (4) and (5):

$$q_u^{I,B,pl} = \frac{16\, B_y}{L^2}\left(1 + \frac{\psi}{8}\right), \qquad \text{(intact frame)}, \tag{32}$$

$$q_c^{B,pl}(n_{r,c}) = \frac{4\, B_y}{L^2(n_{r,c})}\left(1 + \frac{\psi}{4}\right), \qquad \text{(damaged frame)}. \tag{33}$$

where $0 \leq \psi \leq 4$ is a dimensionless parameter depending on the kinematics of the failure mechanism and the beam's slenderness (Masoero et al., 2013). In this section, we briefly investigate the effects of considering catenary actions in the optimal risk-based design of regular frames, using $\psi = 2$.

The catenary limit states derived from Eqs. (32) and (33) correspond to the ultimate limit before complete beam/slab collapses, but substantial (irrecoverable) building damage can be expected before this limit is reached. For the reference frame case, considered in Section 4, reliability indexes corresponding to the "catenary" limit states are shown in Table 2. As observed, catenary effects lead to higher beam reliability indexes. For the damaged condition, in particular, $\beta^{apt}$ shows a significant increase from 2.03 to 3.36. In the risk optimization, this represents a reduced likelihood of ultimate beam collapse; yet, since significant building damage is expected for beam plastic hinge failures, we understand catenary action does not need to be considered in the optimization, as reported in previous sections. The above is also a justification for considering larger failure cost multipliers for column failures, in comparison to beam failures.

For completeness, we briefly report what happens when the additional beam strength, provided by catenary action, is considered in the risk-based optimization. Briefly, optimal design factors for beams, reported in Figures 4, 7, 11 and 12, become significantly smaller, mostly smaller than 0.5. This clearly has no practical significance, as minimal beam strength would most certainly be determined by serviceability (displacement) limit states. Moreover, threshold columns loss probabilities determined as "zero-crossings" of the catenary beam reliability indexes go asymptotically to zero (when compared with Figure 6). Loosely, this means that beam strengthening to produce catenary action has positive cost-benefit for any initial damage probability; an observation which matches practical design recommendations.

## 7. CONCLUDING REMARKS

In this paper, we addressed optimal design of regular frame structures subject to local damage due to abnormal events, leading to loss of beams and columns. We employed a risk-based formulation, which balances the costs of strengthening with expected costs of progressive failure. We employed a simple analytical mechanical model, where beams fail by plastic hinge mechanisms, and columns fail by crushing under compressive loads. The model is limited to gravitational loads.

The analysis covered regular RC frames of different aspect ratios, failure consequences, cost of strengthening and extent of initial damage. Results show that the economic benefit of strengthening frames to bridge over failed load bearing elements (alternative path method or APM) is strongly dependent on threat probabilities. However, threshold local damage probabilities, above which APM design is justified, also depend strongly on frame aspect ratio, consequences of progressive collapse, and cost and reach of the strengthening measures. Typically, APM design is justified for larger threats, taller frames, larger progressive collapse consequences, cheaper strengthening, and limited strengthening measures. Typically, only targeted strengthening actions are cost-effective.

The analysis of optimal design factors for beams and columns, of frames of different aspect ratios, revealed that local bending, local pancake and global pancake failure modes "compete" for the strengthening resources. These resources need to be compensated by effective reductions in expected costs of failure. The optimal allocation of strengthening, between beams and columns, changes according to frame aspect ratio. For lower frames, smaller design margins for beam bending are accompanied by larger margins against column crushing, since bending failures progress upwards, whereas local pancake progresses horizontally. For taller frames, and for greater initial damage, stronger beams need to the accompanied by stronger columns, in order to avoid local and global pancake failures.

Results presented herein were obtained for simple mechanical modelling, but provide useful insight for the optimal cost-benefit analysis and design of realistic structures.

## ACKNOWLEDGEMENTS

Funding of this research project by Brazilian agencies CAPES (Brazilian Higher Education Council), CNPq (Brazilian National Council for Research, grant n. 309107/2020-2) and joint FAPESP-ANID (São Paulo State Foundation for Research - Chilean National Agency for Research and Development, grant n. 2019/13080-9) is also acknowledged.

Valuable comments by the anonymous reviewers are also cheerfully acknowledged.

## CONFLICT OF INTEREST

The author of this manuscript states that there is no conflict of interest.

## DATA AVAILABILITY STATEMENT

The Mathematica algorithms and routines used to obtain results presented in this paper will be made available upon reasonable request.

**TABLES**

Table 1: Random variable statistics.

| Variable | Mean ($\mu$) | C.O.V. ($\sigma/\mu$) | Distribution | Reference |
|---|---|---|---|---|
| Plastic moment strength, bending of RC beams/slabs ($R_{BS}$) | 1.22 | 0.165 | Normal | Beck *et al.* (2020), based on Nowak et al. (2011). |
| Crushing strength of RC columns ($R_C$) | 1.20 | 0.184 | Normal | Beck *et al.* (2020), based on Nowak et al. (2011). |
| Dead load ($D$) | 1.05 $D_n$ | 0.10 | Normal | Ellingwood *et al.* (1980), Ellingwood and Galambos (1982). |
| Live load, arbitrary point in time value ($L_{apt}$) | 0.25 $L_n$ | 0.55 | Gamma | Ellingwood *et al.* (1980), Ellingwood and Galambos (1982). |
| Live load, 50 year ($L_{50}$) | 1.0 $L_n$ | 0.25 | Gumbel | Ellingwood *et al.* (1980), Ellingwood and Galambos (1982). |

Table 2: Reliability index values for *reference* frame.

| | | Intact frame: | | Conditional on initial damage: | |
|---|---|---|---|---|---|
| | | NLC | Strengthened | Damaged | Optimized |
| $\beta^{apt}$ | Global pancake | 3.56 | 3.82 | **3.46** | **3.93** |
| | Local pancake | - | - | **1.80** | **2.62** |
| | Bending | 3,99 | 5,10 | **2.03** | **1.61** |
| | Catenary ($\psi = 2$) | 4.42 | 5.31 | **3.36** | **1.81** |
| $\beta^{50}$ | Global pancake | **2.46** | **2.85** | 2.30 | 3.03 |
| | Local pancake | - | - | -0.02 | 1.08 |
| | Bending | **2.76** | **4.50** | 0.06 | -0.45 |
| | Catenary ($\psi = 2$) | **3.43** | **4.83** | 1.84 | -0.21 |

Table 3: Variants of aspect ratio (number of stories × bays).

| Aspect ratio variations | Number of stories × bays ($n_s \times (n_c - 1)$) | area |
|---|---|---|
| **Tall frame** | (16 × 4) | 64 |
| Intermediate | (13 × 5) | 65 |
| Intermediate | (11 × 6) | 66 |
| *Reference* case: "square" frame | (8 × 8) | 64 |
| Intermediate | (6 × 11) | 66 |
| Intermediate | (5 × 13) | 65 |
| **Low frame** | (4 × 16) | 64 |

Table 4: Other problem variants.

| Set | Problem variants | Case |
|---|---|---|
| 1 | Aspect ratio of individual bays | $L = 2H$ *reference case* <br> $L = H$ with half the tributary area <br> $L = 2H$ by doubling number of columns ($n_c$) <br> $L = 3H$ with 50% increase of tributary area |
| 2 | Cost multipliers | $k_{ductile} = 20$, $k_{brittle} = 40$ different cost multipliers, *reference case* <br> $k_{ductile} = k_{brittle} = 40$ same cost mult. for ductile and brittle failures <br> $k_{ductile} = 40$, $k_{brittle} = 80$ different cost multipliers, increased <br> $k_{ductile} = 50$, $k_{brittle} = 200$ dif. cost multipliers, increased proportion |
| 3 | Strengthening cost | $\alpha_B = \alpha_C = 0.7$ *reference case* <br> $\alpha_B = \alpha_C = 0.9$ higher cost of strengthening <br> $\alpha_B = 0.5, \alpha_C = 0.9$ different cost of strength. for beams/columns <br> $\alpha_B = \alpha_C = 0.7$, but all stories are strengthened ($n_{reinf,s} = n_s$) |
| 4 | Extent of initial damage | $(n_{r,c}^0 \times n_{r,s}^0) = (1 \times 1)$ *reference case* <br> $(n_{r,c}^0 \times n_{r,s}^0) = (1 \times 0)$ reduced extent of initial damage <br> $(n_{r,c}^0 \times n_{r,s}^0) = (2 \times 1)$ increased extent of initial damage <br> $(n_{r,c}^0 \times n_{r,s}^0) = (3 \times 2)$ largest extent of initial damage |

Table 5: Strengthening factors for beams ($B_{sf}$) and columns ($R_{sf}$) for different frames and extents of initial damage (see Eq. (13)).

| Frame | | Initial Damage ($n_{r,c}^0 \times n_{r,s}^0$) | | | |
|---|---|---|---|---|---|
| ($n_s \times (n_c - 1)$) | | $(1 \times 1)$ | $(1 \times 0)$ | $(2 \times 1)$ | $(3 \times 2)$ |
| $(16 \times 4)$ | | 1.38 | 1.42 | 1.98 | 2.47 |
| $(13 \times 5)$ | | 1.29 | 1.34 | 1.87 | 2.29 |
| $(11 \times 6)$ | | 1.24 | 1.29 | 1.78 | 2.17 |
| $(8 \times 8)$ | $R_{sf}$ | 1.15 | 1.23 | 1.66 | 1.95 |
| $(6 \times 11)$ | | 1.08 | 1.17 | 1.55 | 1.74 |
| $(5 \times 13)$ | | 1.04 | 1.15 | 1.48 | 1.60 |
| $(4 \times 16)$ | | 1.00 | 1.13 | 1.40 | 1.40 |
| All | $B_{sf}$ | 2.06 | 2.06 | 4.13 | 6.19 |

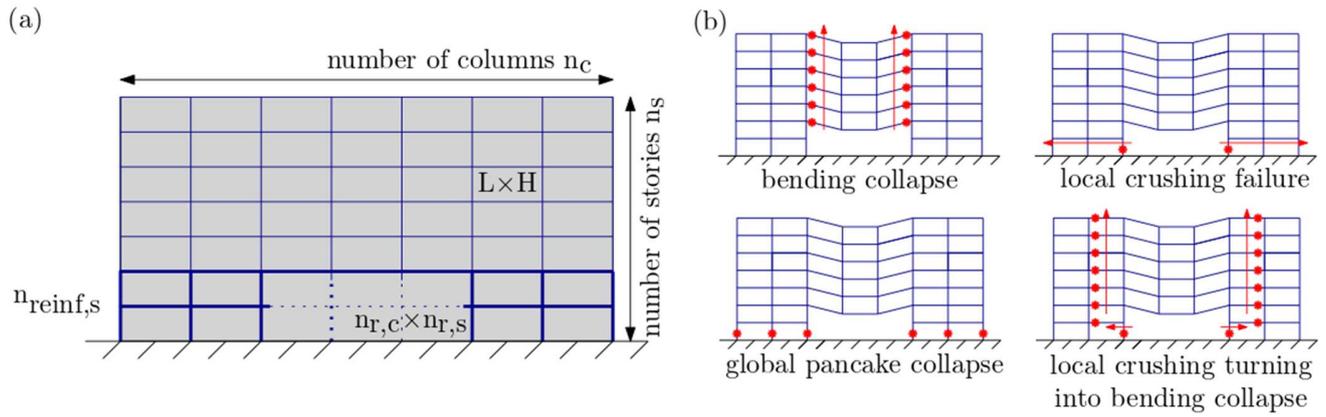

Figure 1: Sketch of regular plane frame (a) and collapse mechanisms (b).

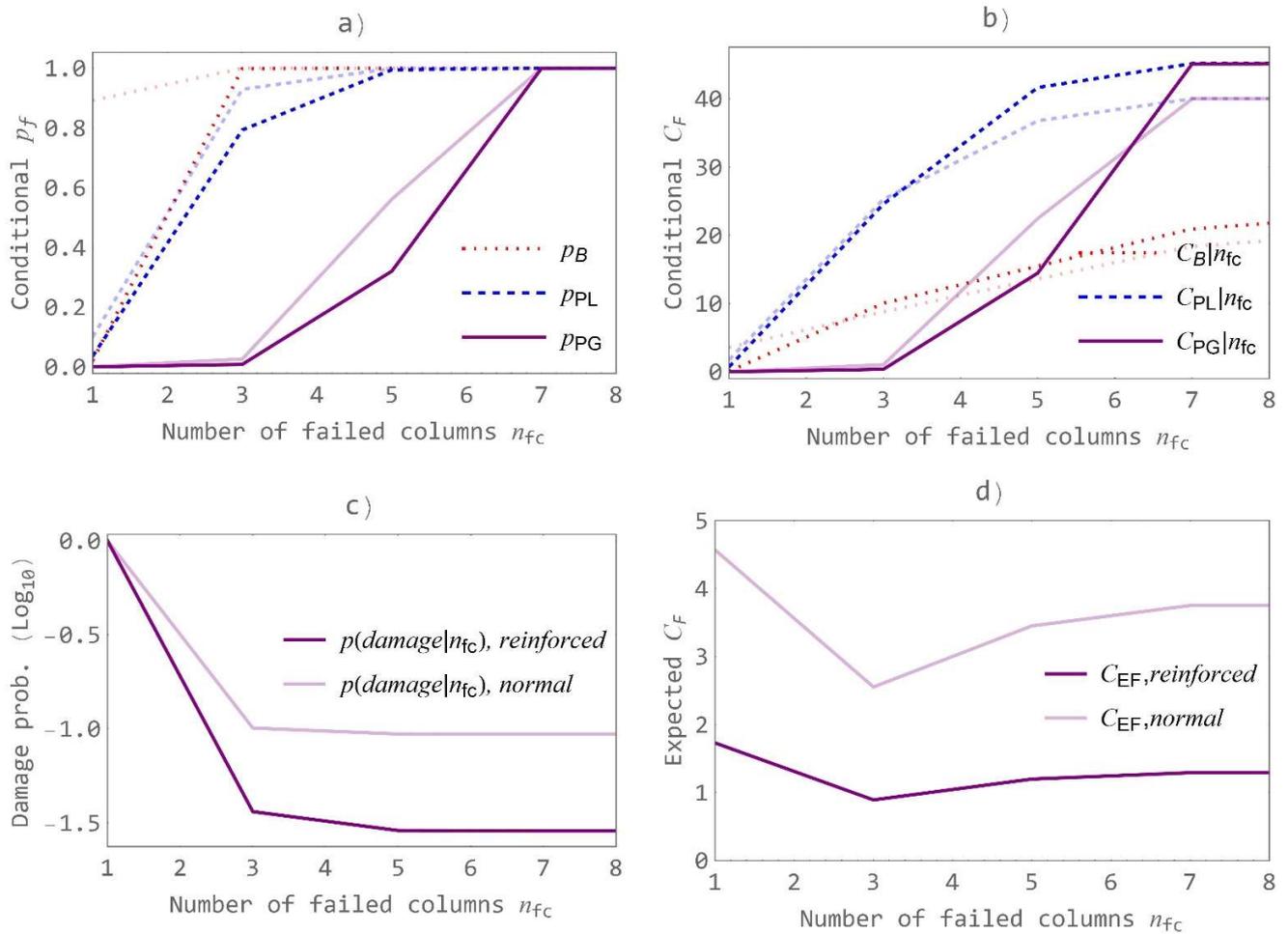

Figure 2: Conditional failure probabilities (a), conditional costs of failure (b), progressive damage probability (c) and expected costs of failure (d) of *normal* (fading lines) and *strengthened* (strong lines) frames, in terms of number of failed columns ($n_{f,c}$).

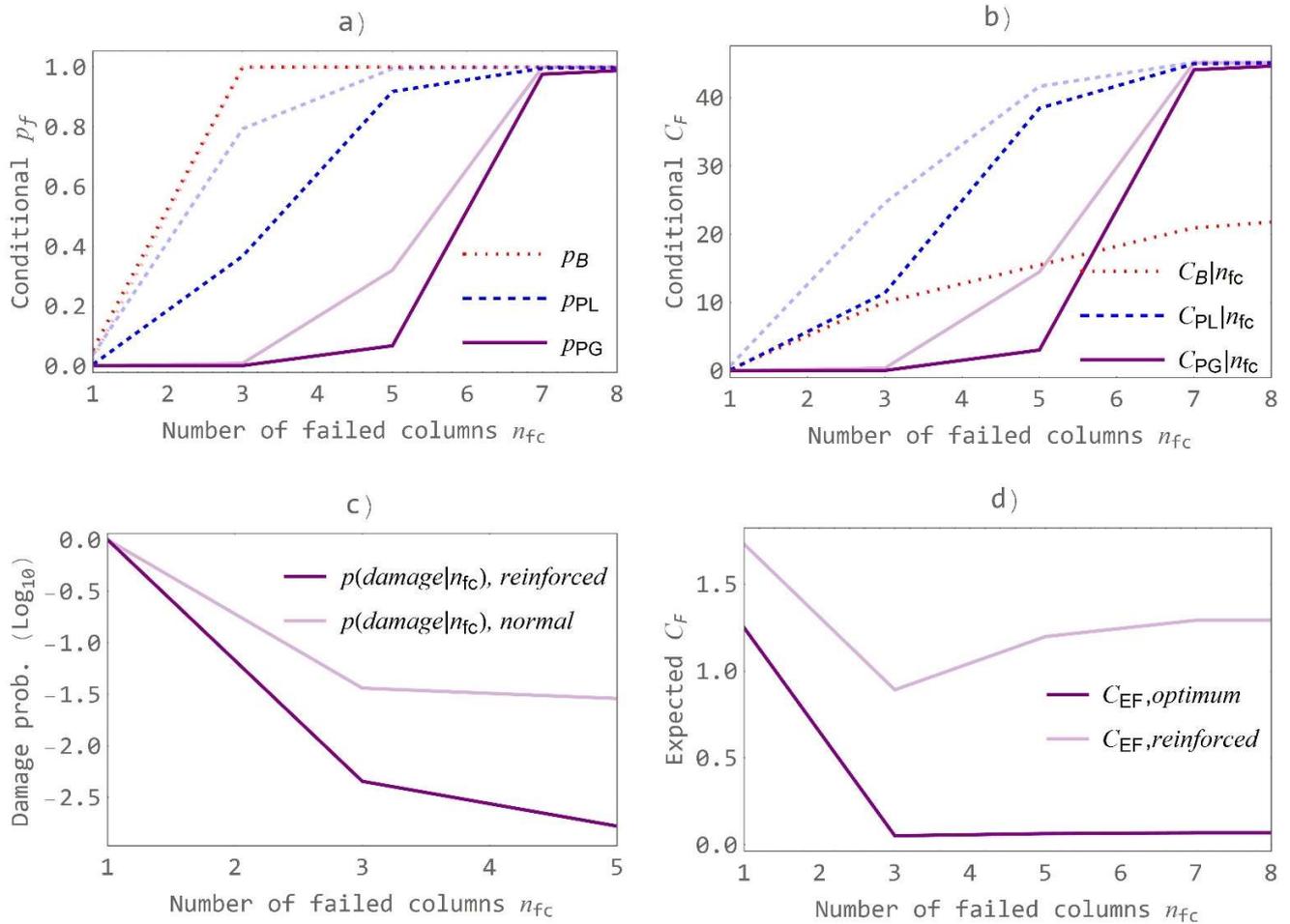

figure 3: Conditional failure probabilities (a), conditional costs of failure (b), progressive damage probability (c) and expected costs of failure (d) of *strengthened* (fading lines, $\lambda_B = \lambda_C = 1$) and *optimally strengthened* (strong lines, $\lambda_B^*$ and $\lambda_C^*$) frames, in terms of number of failed columns ($n_{f,c}$).

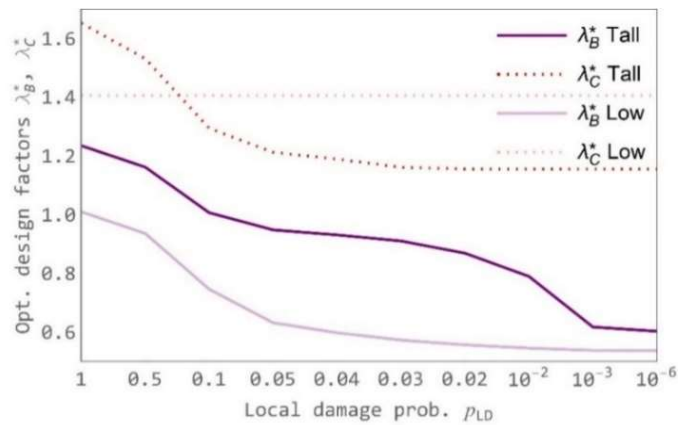

Figure 4: Optimal design factors $\lambda_B^*$ and $\lambda_C^*$ as function of local damage probability (non-linear horizontal scale).

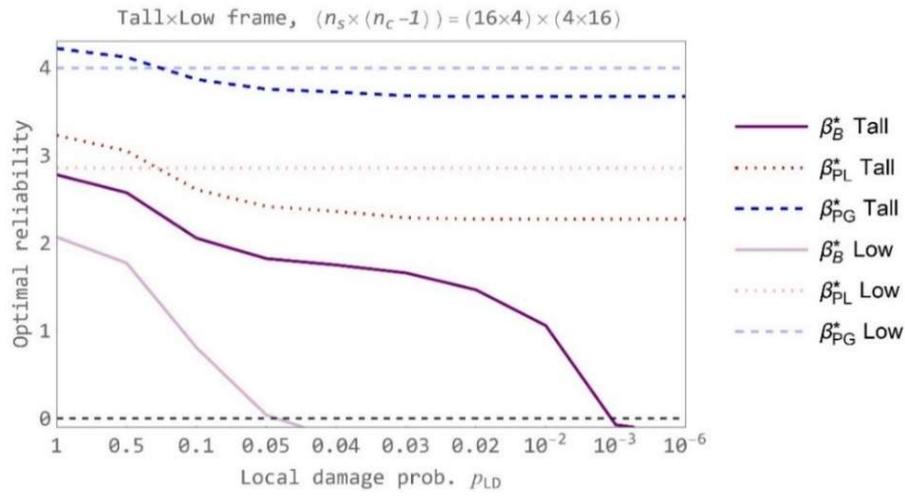

Figure 5: Optimal reliability indexes for bending ($\beta_B^*$), local ($\beta_{PL}^*$) and global pancake ($\beta_{PG}^*$), as function of local damage probability (non-linear horizontal scale).

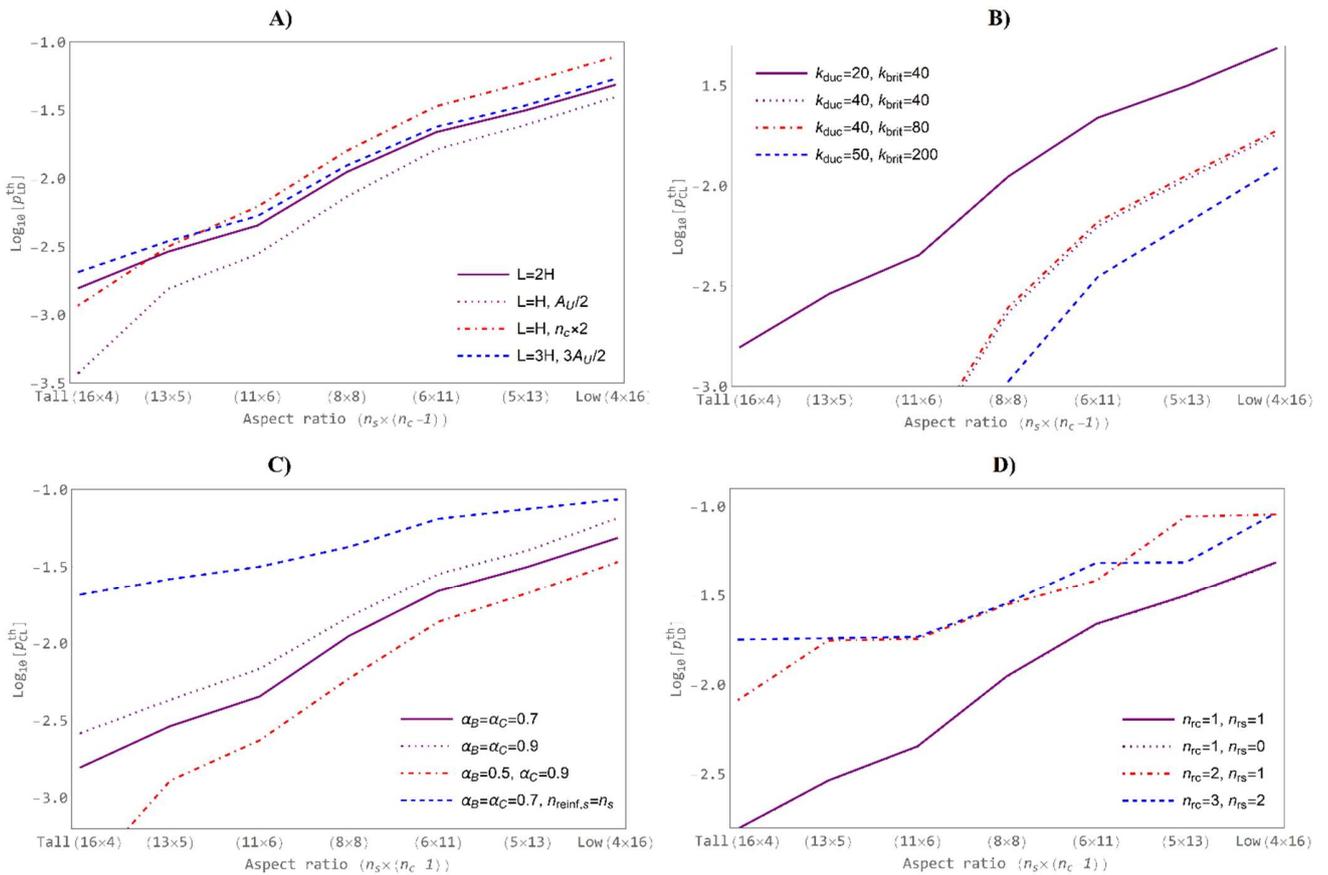

Figure 6: Threshold local damage probability $p_{LD}^{th}$, as function of frame and bay aspect ratios (a), failure cost multipliers (b), costs of strengthening (c) and extents of initial damage.

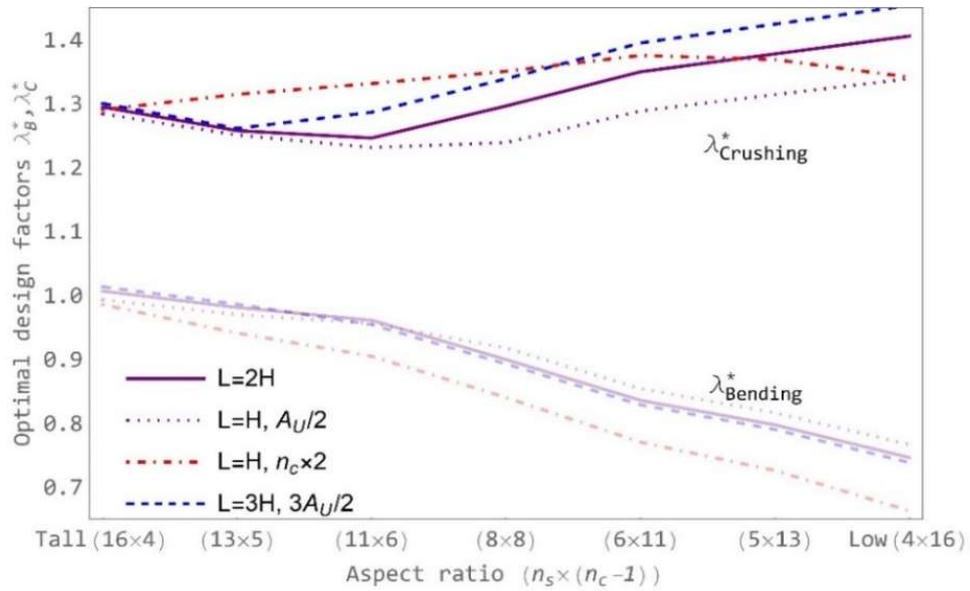

Figure 7: Optimal design factors $\lambda_B^*$ (fading lines) and $\lambda_C^*$ (strong lines), as function of frame and bay aspect ratios, for $p_{LD} = 0.1$.

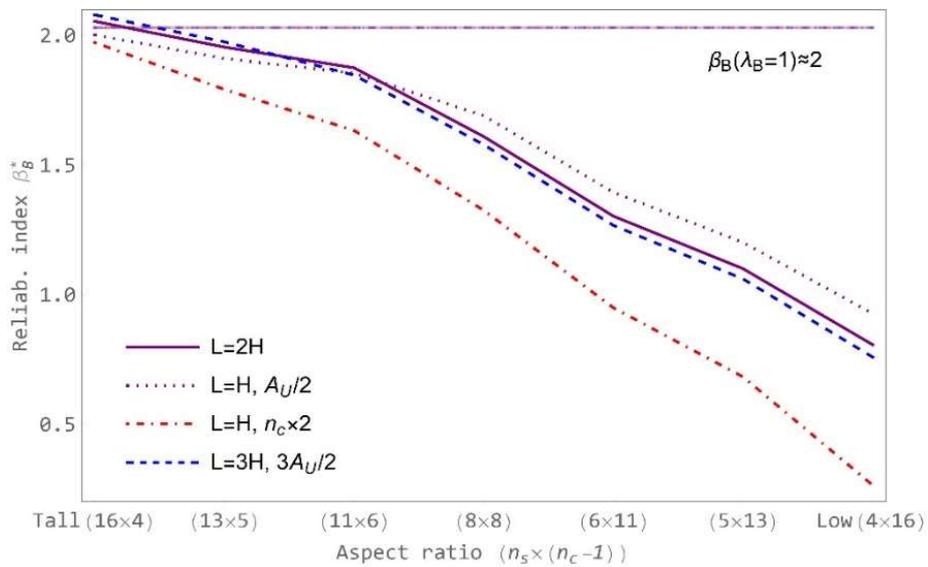

Figure 8: Optimal reliability indexes for bending ($\beta_B^*$), for $p_{LD} = 0.1$.

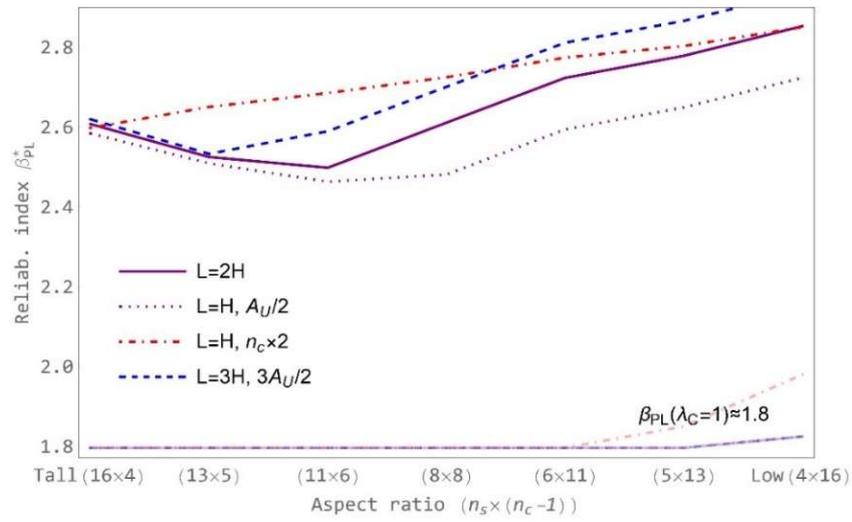

Figure 9: Optimal reliability indexes for local pancake ($\beta^*_{PL}$), for $p_{LD} = 0.1$.

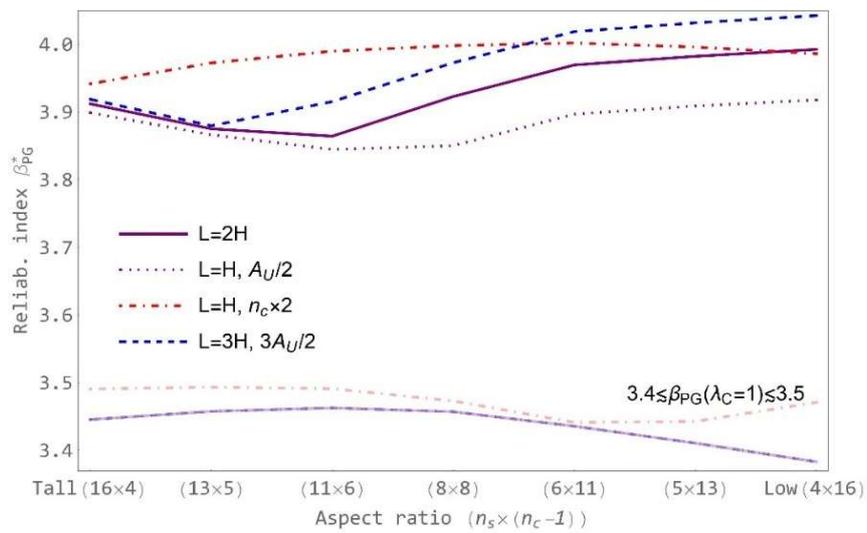

Figure 10: Optimal reliability indexes for global pancake ($\beta^*_{PG}$), for $p_{LD} = 0.1$.

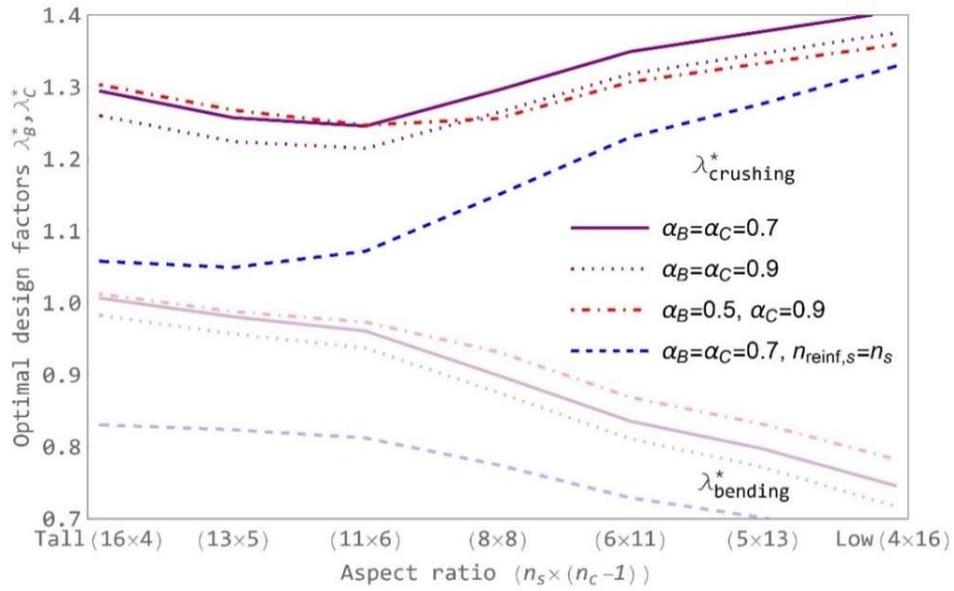

Figure 11: Optimal design factors $\lambda_B^*$ (fading lines) and $\lambda_C^*$ (strong lines), as function of frame aspect ratio, for different costs of strengthening and $p_{LD} = 0.1$.

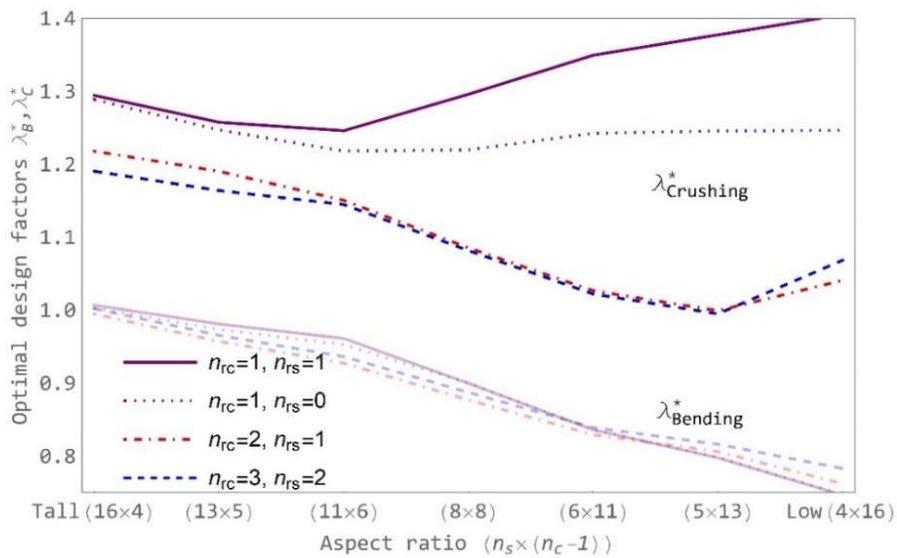

Figure 12: Optimal design factors $\lambda_B^*$ (fading lines) and $\lambda_C^*$ (strong lines), as function of frame aspect ratio, for different extents of initial damage and $p_{LD} = 0.1$.